# FAIR Knowledge Graphs with Semantic Units—a Prototype

Vogt, Lars[1], [1] *TIB Leibniz Information Centre for Science and Technology, Welfengarten 1B, 30167 Hanover, Germany*

orcid.org/0000-0002-8280-0487

Corresponding Author: lars.m.vogt@gmail.com

2# Abstract

Knowledge graphs and ontologies are becoming increasingly important in the context of making data and metadata findable, accessible, interoperable, and reusable (FAIR). We introduce the concept of Semantic Units for organizing Knowledge Graphs into identifiable and semantically meaningful subgraphs. Each Semantic Unit is represented in the graph by its own resource that instantiates a Semantic Unit class. Different types of Semantic Units are distinguished, and together they can organize a Knowledge Graph into different levels of representational granularity with partially overlapping, partially enclosed subgraphs that users of Knowledge Graphs can refer to for making statements about statements. The use of Semantic Units in Knowledge Graphs supports making them FAIR and increases the human-reader-actionability of their data and metadata by increasing the graph's cognitive interoperability by increasing its explorability for a human reader. We introduce a minimal prototype web application for a user-driven FAIR Knowledge Graph that is based on Semantic Units.

# Introduction

The ever-increasing amounts of data being created every day [1–3] require data and metadata to be readily Findable, Accessible, Interoperable, and Reusable (FAIR) [4] for machines and humans alike. Knowledge graphs (KGs) and ontologies, when consistently applying adequate semantic data models, provide a suitable framework for managing and representing FAIR data and metadata [5–10], but also bring their own technical, conceptual, and societal challenges.

One such challenge, which is particularly relevant to user-driven FAIR KGs, involves the question of how to best model statements about statements. What a user considers to be a proposition (=statement) often requires more than one triple of Subject—Predicate—Object when being modelled in a KG (e.g., Anne gives Bob a book). Making statements about statements thus requires being able to reference subgraphs consisting of several triples. While this can be achieved using, e.g., Named Graphs or RDF-Star [11–13], it presupposes knowledge of which triples form a proposition.

Somewhat related to this challenge is the fact that machines need more information than human readers for data and metadata to be actionable, resulting in graphs that are larger and more complex than necessary for human readers. Unlike machines, by considering the context, human readers can often understand the intended meaning of a statement despite missing information, and they also communicate more efficiently than machines by using metaphors, metonymies, and general figures of thought [14]. Consequently, KGs are dealing with the dilemma that the more we push data representations toward machine-actionability, the more complex and therefore the less human-readable the graphs become. This challenge can be solved by decoupling data storage from data representation and showing only data relevant to a human reader in the User Interface (UI). Shape constraint languages such as [SHACL](#) or tools based on them, like [Reasonable Ontology Templates](#) (OTTR) or [LinkML](#), support such decoupling.

However, independent of these challenges, we believe that to increase the general acceptance of FAIR KGs within scientific communities, their cognitive interoperability needs to be increased. This can be achieved by organizing and structuring the graph itself in such ways that tools can support users in exploring and navigating it directly and in meaningful and intuitive ways, zooming in and out of different levels of representational granularity and filtering the graph's contents to what interests them.

In this paper, we describe the concept of Semantic Units [30] with which a FAIR KG can be organized and structured to increase its cognitive interoperability, and we introduce a minimal prototype web application that serves as a proof of concept for Semantic Units.

# Semantic Units

Semantic Units (SUs) [30] structure a KG into identifiable subgraphs, i.e., units of representation that are semantically meaningful to a human reader. Each SU is represented in the KG with its own resource and thus a new type of representational entity [15] in addition to instances and classes. This adds the SUs-graph layer to the data-graph layer of a KG. Each SU resource identifies a particular SU and with it also its corresponding subgraph and thus represents a higher level of abstraction than an individual triple.

Each SU resource instantiates a corresponding SU class. The class includes a human-readable description of the type of information its instances cover. SUs can be organized like a nanopublication [16–18] using Named Graphs, and when properly implemented, they can also function as FAIR Digital Objects [19–21]. The following categories of SUs can be distinguished:

A **Statement Unit** represents the smallest, independent proposition that is semantically meaningful for a human reader. Statement Units mathematically partition the data-graph layer of a KG so that each of its triples belongs to exactly one Statement Unit. Each Statement Unit possesses a subject resource, which is also the subject of its underlying proposition, and, depending on the type of proposition it is modelling, one or more objects. For each type of proposition (e.g., A has-part B; A gives B to C; A travels on the datum B from C to D) exists a corresponding Statement Unit class. Moreover, depending on whether the subject of the proposition refers to a named individual, some unspecified (number of) instance(s) of a class, or all instances of a class, a Statement Unit can be an Assertional, a Contingent, or a Universal Statement Unit.

A **Compound Unit** is a SU that represents a semantically meaningful collection of Statement Units or other Compound Units. The union of the data-graphs of its associated SUs forms its data-graph. Different types of Compound Units can be distinguished:

An **Item Unit** has all Statement Units associated that share the same subject resource, which is also the subject of the Item Unit. Depending on that subject resource, one can differentiate between Instance and Class Item Units.

An **Item Group Unit** associates at least two Item Units that are semantically related to one another through at least one Statement Unit that shares its subject as the subject of one Item Unit and its objects as the subject of another Item Unit.

A **Dataset Unit** is an arbitrary collection of particular SUs and allows users of a user-driven KG to define their own data sets and make statements about them.

A **Granularity Tree Unit** associates two or more Statement Units that in combination represent a granularity tree [22–24], in which resources are related to one another in a tree-hierarchy. Any type of Statement Unit that is based on a partial order relation (e.g., parthood, class-subclass, derives from) can give rise to a granularity perspective [25–27], of which a particular granularity tree is an instance. The resource at the root of the tree is the subject of the Granularity Tree Unit.

A **Granular Item Group** Unit is a Granularity Tree Unit for which for each resource in the tree the corresponding Item Group is associated as well.

## Prototype of a Web KG application that uses Semantic Units

We developed a minimal prototype for a KG application that utilizes some SU classes. The application gets its contents from user input through a web interface, and automatically organizes the resulting KG into partially overlapping and partially enclosed subgraphs, each belonging to a particular SU. The prototype allows users to document a selection of different types of assertions from scholarly publications. In the resulting scholarly KG, each paper is represented as an Item Group Unit (called publication entry unit), with its individual assertions covered by 21 different Statement Unit classes that are organized into 5 different Item Unit classes and one Granularity Tree Unit class.

The application engine is written in Python and uses a local instance of the property graph database [Neo4j](#) as its database. The storage patterns for the different SU classes are specified as a set of [Cypher queries](#) that enable CRUD operations, guaranteeing that all instances of the same SU class are modelled in the same way in the resulting KG, following established graph patterns from the OBO Foundry. Consequently, the data of the KG is FAIR. We also specified a display template for each SU class using Jinja2. Moreover, the prototype includes versioning with a detailed editing history and automatic provenance tracking. The prototype is openly available through https://github.com/LarsVogt/Knowledge-Graph-Building-Blocks.

After initiating the application graph, the application's UI allows you to add a new publication entry by its DOI. Bibliographic metadata from CrossRef is fetched, and the entry is created, i.e., an Item Group Unit with all its required associated Item and Statement Units. Based on a somewhat simplified version of the basic data model for scholarly knowledge discussed in [28], it models a scholarly publication as a report of a research activity, which in turn has some specific research results as its output and realizes some research methods and achieves some objectives. Like a Matryoshka, the Russian stacking doll, this pattern can be applied at finer levels of granularity by allowing to specify additional activities as parts of the overall research activity, and each such added activity can be described using the same pattern, allowing the specification of further activity steps etc.

When a user accesses the data belonging to a paper, the application creates a complex data structure for its respective Item Group Unit and organizes its content using the Jinja templates associated with the types of SUs involved, for displaying it in a web browser, with Item Unit templates having embedded the templates of their associated SUs. The resulting HTML representation of the Item Group Unit is shown in Figure 1. Notice, how the navigation tree to the left represents all Item Units currently associated with the paper entry as a hierarchical tree, from which the users can select the Item Unit they are interested in, and its content is then displayed to the right.

**Fig. 1.** UI showing a scholarly publication item group unit. The left widget shows a navigation tree with the Item Group Unit at the root (i.e., top) and all interlinked Item Units associated with it organized below in a tree. The content of the Item Unit selected in the tree is displayed in the widget to the right, using the corresponding Jinja templates. Here, the Item Unit describing *infectious agent population* (IDO:0000513) is selected that has two connected Statement Units associated that specify the population's basic reproduction number measured to be 2.2 with a 95% confidence interval of 1.9 to 2.6. Input fields allow specifying additional measurements for this quality, specifying parts of the *infectious agent population*, and specifying other qualities.

The underlying graph model allows specifying parts of the research activity reported in the paper as well as parts of its results, of the methods applied, and the objectives achieved. A user can select the Item Unit research result in the navigation tree, and the application provides an input form that allows to add more content (=Statement Units), for example, specifying a type of material entity that the result is about. This creates a material entity Item Unit, and its corresponding template specifies, which types of Statement Units can be added through the UI to further describe it. For a material entity Item Unit, the prototype allows adding has-part and quality-specification statements. Adding another part will not only add the corresponding has-part Statement Unit, but also a corresponding material entity Item Unit for the part and thus allows describing entire partonomies of parts and subparts. In case the added quality is a basic reproduction number (OMIT:0024604) or a weight (PATO:0000128), a basic reproduction number measurement with a 95% confidence interval or a weight measurement Statement Unit can be added.

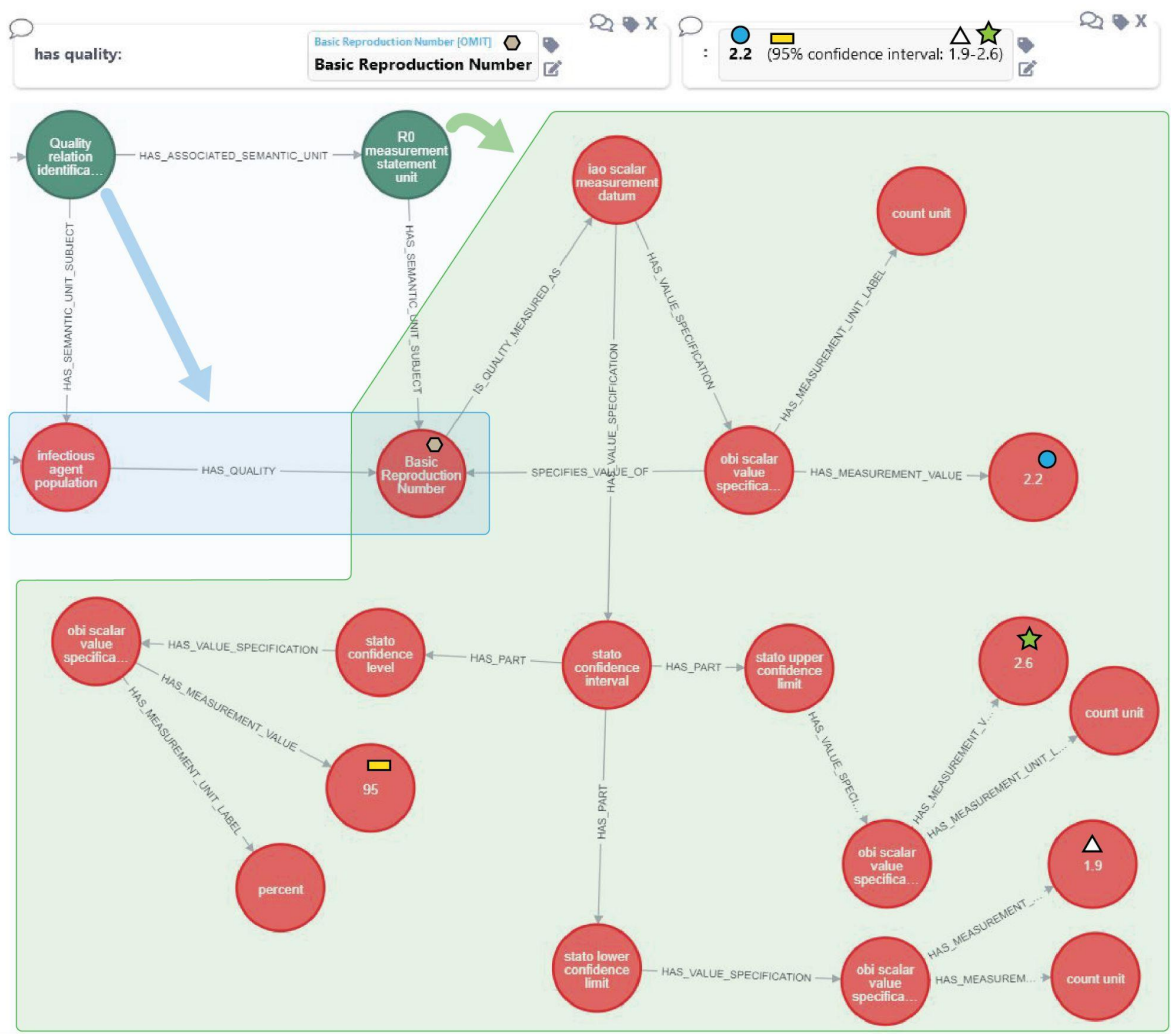

**Fig. 2.** Information in the UI compared to the underlying data-graph for a basic reproduction number measurement. *Top*: The UI displays the information from the quality and the basic reproduction number (R0) measurement Statement Unit in a single line, indicating that the respective population has an R0 value of 2.2, with a 95% confidence interval of 1.9 to 2.6. It uses information from only five nodes of the underlying data-graph (indicated by colored symbols). *Bottom*: The same information, as it is stored in the Neo4j database of the application. The red nodes and their relations represent the contents of the two Statement Units (green nodes) — the data-graph of the quality Statement Unit shown in the blue box and the data-graph of the R0 measurement Statement Unit in the green box.

The application identifies when at least two parts have been added and adds a parthood Granularity Tree Unit that includes all has-part Statement Units that are connected to each other, forming a partonomy. A corresponding element is added to the navigation tree, and the application allows viewing the corresponding partonomy graph.

Although the knowledge graph is user-driven, all data are FAIR, as the application uses controlled vocabularies wherever appropriate through the [BioPortal REST API](#) for autocomplete input fields, and the same type of data is modelled using the same storage model by using the same generic CREATE/MERGE Cypher queries. The resulting graph models statements by relating newly created instances of respective ontology classes to one another in an ABox graph, with each node and relation specifying a corresponding URI.

The prototype also demonstrates how human-readable data display is decoupled from machine-actionable data storage. After having added the basic reproduction number measurement with the 95% confidence interval, the information is displayed in the UI as a single line of information, whereas in the graph it comprises 20 triples organized into two Statement Units (Fig. 2).

For each statement unit, users can specify the level of the statement's certainty or confidence. This is a statement about a statement, linking the node of the target Statement Unit to a resource indicating the certainty of the assertion — information that can be useful for identifying citation distortions, in which mere possible claims become well-established facts through a chain of citations [29].

Because the application implements a soft-delete and a versioning approach, it is possible to provide a detailed editing history for each SU and each of its input-fields. Moreover, users can make snapshots of the current data- and SUs-graph of a given SU along with all information coming from its associated SUs. This creates a distinct resource with its own URI representing the version. All nodes and relations belonging to a specific version are identifiable and can be referenced, making it a citable resource that is accessible independent of subsequent editing steps.

Also note that when adding content through the UI, users do not have to think about how to model the data in a semantic way using specific data schemata or graph patterns, since these are provided by pre-defined Cypher queries for the respective SU classes.

## Conclusion

Each SU specifies the graph pattern used for semantically modelling its contents, thus providing a standard for ABox patterning, increasing the FAIRness of the KG. The Jinja templates decouple data storage from data display, showing only relevant contents in the UI. Also, making statements about statements is straightforward, since the graph is already organized into semantically meaningful subgraphs, each of which can be referenced by the corresponding SU resource. It is time to develop new tools for exploring the graph, thereby increasing its cognitive interoperability. Besides machine-actionability, we must also focus on the human-reader-actionability of data and metadata and turn FAIR KGs into FAIRer knowledge graphs. SUs, with the additional layers of abstraction they add to a KG, is one suggestion how we could arrive there.